**Characterizations of the Nonlinear Optical Properties for (010) and ($\bar{2}$01) Beta-phase Gallium Oxide**

*Hong Chen, Houqiang Fu, Xuanqi Huang, Jossue A. Montes, Tsung-Han Yang, Izak Baranowski and Yuji Zhao\**

*School of Electrical, Computer and Energy Engineering, Arizona State University, Tempe, AZ 85287, U.S.A.*

*\*yuji.zhao@asu.edu*

**Abstract:** We report, for the first time, the characterizations on optical nonlinearities of beta-phase gallium oxide ($\beta$-Ga$_2$O$_3$), where both (010) $\beta$-Ga$_2$O$_3$ and ($\bar{2}$01) $\beta$-Ga$_2$O$_3$ were examined for two-photon absorption (TPA) coefficient, Kerr refractive index, and their polarization dependence. The wavelength dependence of the TPA coefficient and Kerr refractive index was estimated using a widely used analytical model. $\beta$-Ga$_2$O$_3$ exhibits a TPA coefficient of 1.2 cm/GW for (010) $\beta$-Ga$_2$O$_3$ and 0.58 cm/GW for ($\bar{2}$01) $\beta$-Ga$_2$O$_3$. The Kerr refractive index is $-2.14 \times 10^{15}$ cm$^2$/W for (010) $\beta$-Ga$_2$O$_3$ and $-2.89 \times 10^{15}$ cm$^2$/W for ($\bar{2}$01) $\beta$-Ga$_2$O$_3$. In addition, $\beta$-Ga$_2$O$_3$ shows stronger nonlinear optical anisotropy on the ($\bar{2}$01) plane than on the (010) plane, possibly due to highly asymmetric crystal structure. Compared with that of gallium nitride (GaN), the TPA coefficient of $\beta$-Ga$_2$O$_3$ is 20 times smaller, and the Kerr refractive index of $\beta$-Ga$_2$O$_3$ is also found to be 4–5 times smaller. These results indicate that $\beta$-Ga$_2$O$_3$ has the potential for use in ultra-low loss waveguides and ultra-stable resonators and integrated photonics, especially in the UV and visible wavelength spectral range.

## 1. Introduction

As an emerging wide bandgap semiconductor material, beta-phase gallium oxide (β-Ga$_2$O$_3$) has attracted considerable attention in catalysis[1,2], gas sensors[3], power electronics[4] and potential optical devices such as waveguides and resonators. Due to its wide bandgap, β-Ga$_2$O$_3$ possesses a board transparent spectrum from ultraviolet (UV) to visible wavelengths, ideal for optical applications at this wavelength range. Due to the compatibility of β-Ga$_2$O$_3$ with the III-nitride material system[9], β-Ga$_2$O$_3$ optical devices can also be actively-integrated with III-nitride based UV–visible light sources[10–14] and detectors[15–16]. These unique properties make β-Ga$_2$O$_3$ a very promising candidate for emerging integrated photonics applications at elusive UV to visible wavelengths, which are critical for various optical applications such as biochemical sensing, UV Raman spectroscopy, frequency up/down conversion, and quantum emitters, etc[17]. In order to realize the full potential of β-Ga$_2$O$_3$ materials in these optical applications, a comprehensive investigation of the material properties of β-Ga$_2$O$_3$ is of crucial importance. However, previous studies on β-Ga$_2$O$_3$ have been mainly focused on the electronic[5] and thermal properties[9] of β-Ga$_2$O$_3$, focusing on power electronics applications, with only very limited reports on the simple characterizations of basic optical properties on β-Ga$_2$O$_3$, such as transmission[6] and refractive index[7,8]. The fundamental nonlinear optical properties of β-Ga$_2$O$_3$, which are vital for various optical applications such as integrated photonics and quantum photonics, have never been investigated.

Nonlinear optical processes such as the two-photon absorption (TPA) coefficient and the Kerr effect play a significant role in determining the performance of optical devices. For example, the TPA process is one of the major optical loss mechanisms for optical waveguides and resonators under high optical power density[18,19]. For resonators operating by critical coupling from bus waveguides, the refractive index shifting, mainly governed by the Kerr refractive index at high power density, will affect the coupling efficiency especially for ultra-high quality factor resonators[21]. To develop high performance β-Ga$_2$O$_3$ optical devices, it is therefore of paramount importance to characterize and understand the nonlinear optical properties of β-Ga$_2$O$_3$ including the TPA coefficient and Kerr refractive index.

To investigate and evaluate the nonlinear optical properties of β-Ga$_2$O$_3$ in the visible spectral range, we performed a typical Z-scan characterization to study the TPA coefficient and Kerr refractive index of β-Ga$_2$O$_3$. The results show that β-Ga$_2$O$_3$ has a much smaller TPA coefficient (20 times smaller) and Kerr refractive index (4–5 times smaller)[19] compared to GaN, which is ideal for high performance waveguides and resonators in visible and potential UV wavelengths. Furthermore, due to the highly asymmetric crystalline structure of β-Ga$_2$O$_3$, the optical nonlinearity is found to be highly anisotropic. Relatively stronger nonlinear optical anisotropy was observed on the ($\bar{2}$01) plane than on the (010) plane. These results will serve as important references and guidelines for the design and fabrication of future photonic devices based on β-Ga$_2$O$_3$.

This paper is organized as the following: In Section 2, we describe the methods used in this study, including experimental setup and theoretical models. In Section 3, we show the experimental results on β-Ga$_2$O$_3$ materials and discuss their impacts on the photonic devices. In section 4, we provide a brief summary of the work.

## 2. Methods

**Figure 1(a)** schematically shows the experimental setup of the Z-scan measurement[19,21,22] used in this study. Light source was an ultrafast titanium-sapphire laser operating at 808 nm. A $\chi^2$ crystal was used to generate the second harmonic wave at the wavelength of 404 nm. A half wavelength plate working at 404 nm was placed after the $\chi^2$ crystal for light polarization tuning. Light beam was expanded by a set of lenses before being sent into optical objectives in order to fully utilize the numerical aperture. The samples were positioned between two optical objectives. An aperture was implemented in front of a power meter to perform open and closed aperture testing.

The unintentionally doped (UID) β-Ga$_2$O$_3$ samples used in this work were provided by Tamura Corporation with a carrier concentration on the order of ~10$^{17}$ cm$^{-3}$ and a thickness of ~ 500 µm. The backside of the samples was polished by a hand-grinding method with diamond lapping film of 0.5 µm grade. The polishing process was carefully controlled so that the thickness of samples after polishing was reduced by less than 50 µm. **Figure 1(b)** depicts the crystal structure of β-Ga$_2$O$_3$, where the (010) and ($\bar{2}$01) crystal planes are also illustrated.

To test the reliability of the setup, the beam size of the output light is used as an indicator. Based on Gaussian optics, with an incident beam diameter of 4 mm and a numerical aperture (NA) of 0.25, the calculated diameter of the output light is ~5 µm (at 1/e$^2$) as shown in **Figure 1(c)**. This value is consistent with beam diameter we extracted from open aperture Z-scan measurement in next section.

The TPA coefficient and Kerr refractive index can be obtained from open/closed aperture scanning based on Equation 1–3,

$$T \approx 1 - \frac{\alpha_{TPA} I_0 L_{eff}}{2\sqrt{2}} \times \frac{1}{1 + Z^2/Z_0^2} \qquad (1)$$

$$Z_0 = \frac{n\pi\omega_0^2}{\lambda} \qquad (2)$$

$$n_{kerr} = \frac{\lambda \, \Delta\phi}{2\pi I_0 L_{eff}} \qquad (3)$$

where $T$ is the normalized transmission, $I_0$ is the peak beam power density, $L_{eff}$ indicates effective sample length, $Z_0$ indicates Rayleigh range of the beam, $n$ is the refractive index, $\omega_0$ is the beam size at the focal plane, and $\lambda$ is the wavelength. $L_{eff}$ refers to the sample thickness, $\Delta\phi$ the nonlinear phase shift due to Kerr effect[21]. More information about these equations can be found in References 18, 20–24. It should be noted that the samples are tested at 404 nm which is above the half bandgap energy of β-Ga$_2$O$_3$ (506 nm). Therefore, a three-photon absorption modification to Equation 1 is not required in our case[26]. The wavelength-dependence of the TPA coefficients is also theoretically calculated using Equation 4,

$$\alpha_{TPA}(\omega) = K \frac{\sqrt{E_p}}{n_0^2 E_g^3} F_2(\hbar \cdot \omega / E_g) \qquad (4)$$

in which $E_g$ is the direct bandgap energy, $E_p = 2|P_{vc}|^2/m_0$ is a material-independent parameter for direct bandgap semiconductors obtained by the $k \cdot p$ model[27], $n_0$ is the refractive index, $\omega$ is the frequency, and $K$ is a material-independent constant. $F_2$ is a fitting function with the form $F_2(x) = (2x-1)^{1.5}/(2x)^5$. It should be noted that β-Ga$_2$O$_3$ exhibits a direct bandgap energy of 4.9 eV and a slightly smaller indirect bandgap of 4.85 eV[1]. In our theoretical analysis, we do not consider phonon assisted TPA because it only contributes to optical nonlinearity at a narrow bandwidth.

## 3. Experimental results and discussions

**Figure 2** shows the open aperture scanning results for (010) and ($\bar{2}$01) β-Ga$_2$O$_3$. Fitting the experimental data with Equation 1–3 yielded the TPA coefficients and Kerr refractive index, which were summarized in **Table 1**. (010) β-Ga$_2$O$_3$ showed a TPA coefficient $\alpha_{TPA} = 1.2$ cm/GW and $n_{kerr} = -2.14 \times 10^{-15}$ cm$^2$/W when $\vec{E} \perp [102]$, while an $\alpha_{TPA}$ of 0.58 cm/GW and $n_{kerr}$ of $-2.89 \times 10^{-15}$ cm$^2$/W were obtained on ($\bar{2}$01) β-Ga$_2$O$_3$ when $\vec{E} \parallel [102]$. Due to the smaller TPA coefficient on ($\bar{2}$01) β-Ga$_2$O$_3$ than that on (010) β-Ga$_2$O$_3$, optical devices such as waveguides and resonators on ($\bar{2}$01) β-Ga$_2$O$_3$ will have lower optical loss induced by TPA at the visible spectral regime. In comparison, ($\bar{2}$01) β-Ga$_2$O$_3$ showed a larger Kerr refractive index than that on (010) β-Ga$_2$O$_3$. In addition, the output beam diameters were also obtained, which are in good agreement with theoretical calculations based on Gaussian optics. This indicates the setup is reliable and accurate.

There are several physical models that can be implemented to understand this highly anisotropic optical nonlinearity[28–33]. A quantum mechanical approach[28,29] calculates third order susceptibility accurately but requires intensive computing resources. Therefore it is only utilized in simple crystal structures[29]. A simplified bond-orbital model developed in Reference

30–32 successfully explained the optical nonlinearities for various kinds of materials including metal-oxide crystals[33]. We employed this model in this work to qualitatively understand the anisotropic nonlinearity of (010) and ($\bar{2}$01) β-Ga$_2$O$_3$. From the bond-orbital model, the optical anisotropy of a crystal is mainly contributed from bonding electrons between adjacent atoms[30], while for other electrons that screen around individual atoms, the contribution to the optical anisotropy is less significant. As investigated comprehensively in Reference 34, β-Ga$_2$O$_3$ is constructed by two types of gallium ions and three types of oxygen ions as shown in **Figure 3(a) and 3(b)**. Type I gallium ion Ga$_{(I)}$ is surrounded by a distorted tetrahedron of oxygen ions with atomic distances from 1.80 A to 1.85 A (1.83 A in average), type II gallium ion Ga$_{(II)}$ is surrounded by a distorted octahedron with atomic distances from 1.95 A to 2.08 A (2.00 A in average). On the (010) plane of β-Ga$_2$O$_3$, the bonds are not orderly oriented and therefore electrical fields in the (010) plane are more likely to interact with most of the bonds. However, on the ($\bar{2}$01) plane, bonds between Ga$_{(I)}$ and O$_{(III)}$ are highly oriented along the (010) direction. Such bonds have a relatively weak interaction with electric fields polarized in ($\bar{2}$01) plane. The consequence is that the optical nonlinearity on the ($\bar{2}$01) plane is relatively weaker than that on the (010) plane.

We previously estimated the TPA coefficients and Kerr refractive index for GaN[19] at same wavelength range. The TPA coefficient of (010) and ($\bar{2}$01) β-Ga$_2$O$_3$ are ~10 and ~20 times smaller than that of GaN at 404 nm, respectively. The high TPA coefficient observed in GaN might due to its exciton effects[22]. This result implies that the β-Ga$_2$O$_3$ material is more capable of handling high optical power density applications in visible wavelength spectral range. Furthermore, Kerr coefficients of β-Ga$_2$O$_3$ are also 4–5 times smaller than that of GaN. For optical applications that require critical coupling, e.g., coupling from bus waveguide to ring or disk resonators, the Kerr effect modifies the coupling efficiency at high optical power density, especially in high resonance quality factor resonators[20]. Therefore, β-Ga$_2$O$_3$-based resonators will exhibit extreme coupling stability under high power operation compared to GaN-based resonators.

**Figure 4** represents the polarization dependences of the TPA and Kerr coefficients of (010) and ($\bar{2}$01) β-Ga$_2$O$_3$. Since β-Ga$_2$O$_3$ has a monoclinic crystal structure (Figure 1), 41 independent nonzero elements are required to fully describe its third order nonlinearity $d_{il}$. Therefore, it is very difficult and inconvenient to find an explicit expression for $d_{eff}$ using polarization angle and individual nonlinear component $d_{il}$[20–22,25]. To describe the anisotropic nonlinearity clearly, we used $\Delta T_{max}/\Delta T_{min}$ in this work. Relatively higher nonlinear optical anisotropy was found on ($\bar{2}$01) β-Ga$_2$O$_3$ with $\Delta T_{max}/\Delta T_{min}$ of 1.93 while (010) β-Ga$_2$O$_3$ had a $\Delta T_{max}/\Delta T_{min}$ of 1.29.

**Figure 5** shows the wavelength dependence of the TPA coefficient using Equation 4 for (010) and ($\bar{2}$01) β-Ga$_2$O$_3$. For (010) β-Ga$_2$O$_3$, $E\perp$ has larger TPA coefficient than $E\parallel$. For ($\bar{2}$01) β-Ga$_2$O$_3$, $E\perp$ has a smaller TPA coefficient than $E\parallel$. The highest TPA coefficient is observed in (010) β-Ga$_2$O$_3$ when the electric field is perpendicular to the [102] direction, while the lowest TPA coefficient is obtained on ($\bar{2}$01) β-Ga$_2$O$_3$ when the electric field is parallel to the [102] direction. For each plane, the polarization dependence of the TPA coefficient is a function of multiple physical parameters[34] such as bond iconicity, covalent radii, etc. The physical mechanism of this polarization dependence on different crystal orientations of β-Ga$_2$O$_3$ is a topic of on-going investigation. For both ($\bar{2}$01) β-Ga$_2$O$_3$ and (010) β-Ga$_2$O$_3$ samples, the maximum TPA process occurs at the ~360 nm wavelength. For wavelengths above 360 nm, the TPA coefficient decreases as wavelength increases, which can be attributed to the decreased excitation energy with increasing wavelength. For wavelengths below 360 nm, the TPA coefficient increases as the transition approaches its resonance wavelength.

## 4. Conclusions

We characterized the TPA coefficient and Kerr refractive index of both (010) and ($\bar{2}$01) β-Ga$_2$O$_3$. The TPA coefficient of Ga$_2$O$_3$ was found to be 10 to 20 times smaller than that of GaN at 404 nm. The Kerr refractive index of $Ga_2O_3$ was 4 to 5 times lower than that of GaN. Therefore, due to its ultra-low TPA coefficient and its small Kerr refractive index, $β-Ga_2O_3$ has the potential to serve as a more efficient platform for integrated photonic applications in UV and visible spectral range. Furthermore, the optical nonlinearities of $β-Ga_2O_3$ are highly anisotropic due to its asymmetric crystal structure. These results can serve as guidelines for the design and fabrication of $β-Ga_2O_3$-based integrated photonic devices at UV–visible wavelengths for various optical applications such as biochemical sensing, UV Raman spectroscopy, frequency up/down conversion, and quantum photonics.

**Acknowledgements**

The authors thank Dr. Michael Gerhold of US Army Research Office for the helpful discussion. This work is supported by the Bisgrove Scholar Program from Science Foundation Arizona. We gratefully acknowledge the use of facilities within the LeRoy Eyring Center for Solid State Science and CLAS Ultra Fast Laser Facility at Arizona State University.

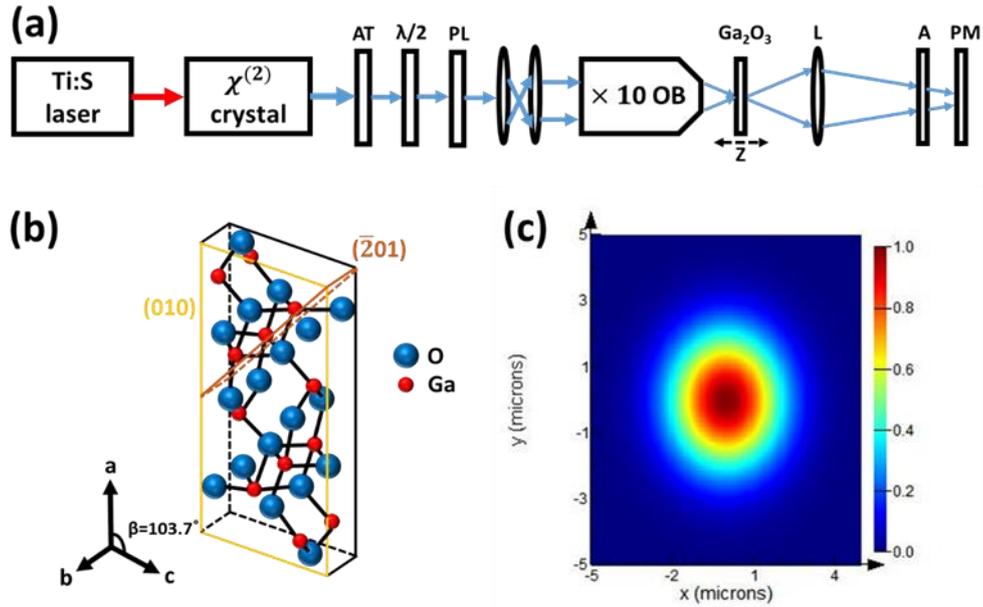

**Figure 1.** (a) Experimental setup of the Z-scan measurement used in this work. AT indicates the attenuator, λ/2 is the half-wavelength plate, PL is the polarizer, OB is the optical objective lens, A is the aperture, PM is the power meter. L1, L2, L3 are lenses. (b) Crystalline structure of β-$Ga_2O_3$ showing (101) and ($\bar{2}$01) crystal orientations. (c) The estimated field intensity at the focal point, which is in good agreement with our open aperture fitting.

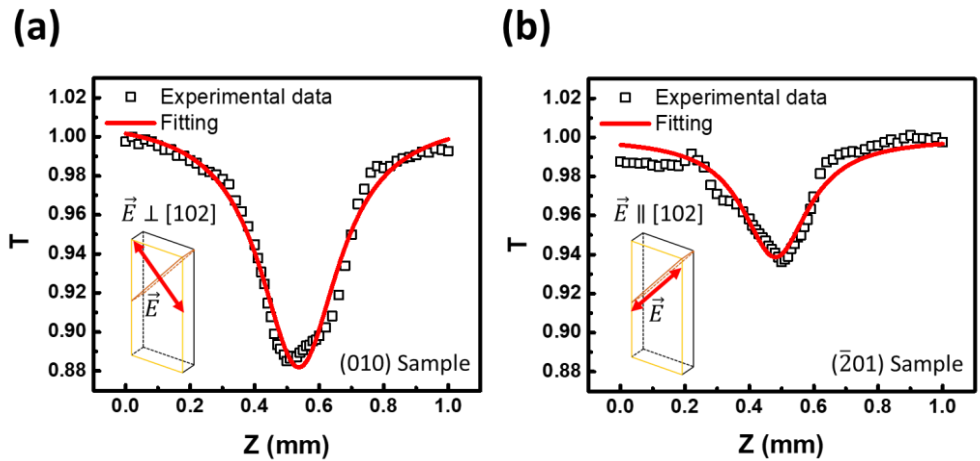

**Figure 2.** Typical open aperture scanning curve obtained in this work. (a) Scan curve for (010) β-$Ga_2O_3$ sample, (b) for ($\bar{2}$01) β-$Ga_2O_3$ sample.

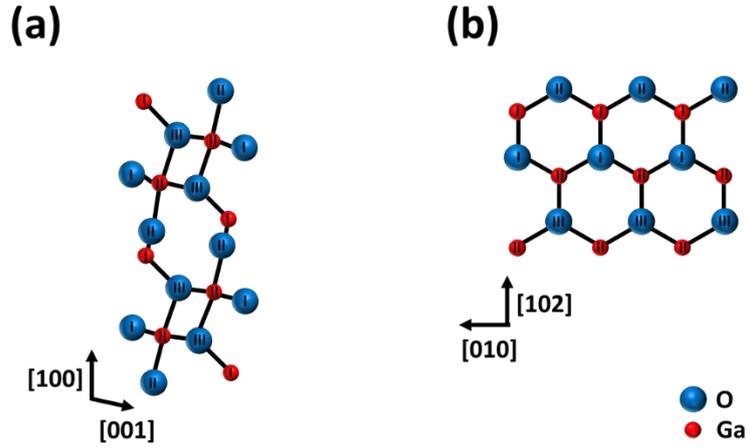

**Figure 3.** Schematic views for the arrangement of ions inside β-Ga$_2$O$_3$, different types of ions are marked out by numbers. (a) Schematic view for (010) β-Ga$_2$O$_3$ sample, (b) for ($\bar{2}$01) β-Ga$_2$O$_3$ sample.

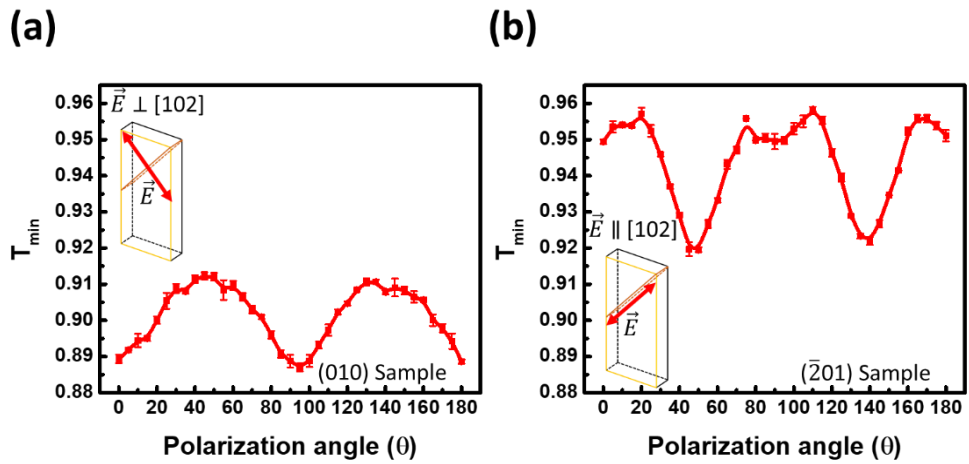

**Figure 4.** Polarization dependence of minimum transmittance obtained in this work. (a) Polarization dependence for (010) β-Ga$_2$O$_3$ sample, (b) for ($\bar{2}$01) β-Ga$_2$O$_3$ sample.

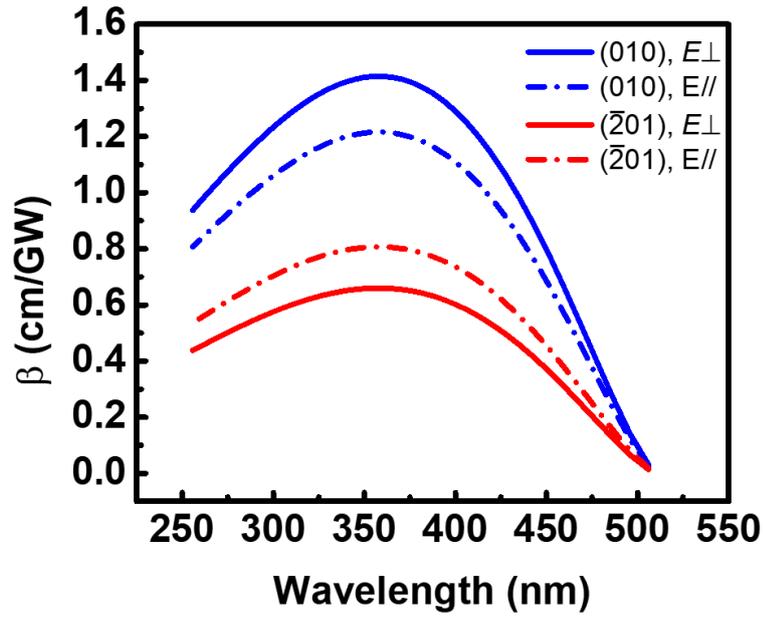

**Figure 5.** The estimated wavelength dependence of the TPA coefficient for (010) and ($\bar{2}$01) β-Ga$_2$O$_3$ samples. "$E\perp$" indicates that the electrical field intensity is perpendicular to [102] direction, while "$E\|$" indicates that field intensity is parallel to [102] direction

**Table 1.** TPA coefficients and Kerr coefficients obtained in this work for (010) and ($\bar{2}$01) β-Ga$_2$O$_3$ samples. Beam diameter from fitting parameter is also shown.

| (010), $\vec{E} \perp [102]$ | | | ($\bar{2}$01), $\vec{E} \| [102]$ | | |
|---|---|---|---|---|---|
| $\alpha_{TPA}$(cm/GW) | $n_{kerr}$(cm$^2$/W) | R(μm) | $\alpha_{TPA}$(cm/GW) | $n_{kerr}$(cm$^2$/W) | R(μm) |
| 1.20 | $-2.14\times10^{-15}$ | 5.8 | 0.58 | $-2.89\times10^{-15}$ | 5.68 |